\documentclass[conference]{IEEEtran}
\IEEEoverridecommandlockouts

\usepackage{cite}
\usepackage{amsmath,amssymb,amsfonts,bbm}
\usepackage{dsfont}
\usepackage{algorithmic}
\usepackage{graphicx}
\usepackage{textcomp}
\usepackage{xcolor}
\usepackage{booktabs}
\usepackage{array}
\usepackage{url}
\usepackage{algorithm}
\usepackage{algorithmic}

\def\BibTeX{{\rm B\kern-.05em{\sc i\kern-.025em b}\kern-.08em
    T\kern-.1667em\lower.7ex\hbox{E}\kern-.125emX}}

\makeatletter
\def\ps@IEEEtitlepagestyle{%
  \def\@oddfoot{\mycopyrightnotice}%
  \def\@evenfoot{}%
}
\def\mycopyrightnotice{%
  {\footnotesize © 2026 IEEE. Personal use of this material is permitted. Permission from IEEE must be obtained for all other uses. Accepted in IEEE GLOBECOM 2026 \hfill}
  \gdef\mycopyrightnotice{}
}
\makeatother

\begin{document}
\bstctlcite{IEEEexample:BSTcontrol} % to shrink citations

% ---- Title -------------------------------------------------------------------
\title{Goal-Oriented Communication and Control Co-Design via Semantic Push-Pull in Industrial IoT}
% ---- Author placeholders -----------------------------------------------------
\author{\IEEEauthorblockN{Muhammad Azeem Khan\IEEEauthorrefmark{1}\IEEEauthorrefmark{2}, Yuriy Zacchia Lun\IEEEauthorrefmark{2}, Aamir Mahmood\IEEEauthorrefmark{1}, Piergiuseppe Di Marco\IEEEauthorrefmark{2}, \\ Mikael Gidlund\IEEEauthorrefmark{1}, and Fortunato Santucci\IEEEauthorrefmark{2} }	\IEEEauthorblockA{\IEEEauthorrefmark{1}Department of Computer and Electrical Engineering, Mid Sweden University, 851 70 Sundsvall, Sweden\\ \IEEEauthorrefmark{2}Department of Information Engineering, Computer Science and Mathematics, University of L’Aquila, 67100 L’Aquila, Italy \\ 
  \IEEEauthorrefmark{1}Email: muhammadazeem.khan@graduate.univaq.it, yuriy.zacchialun@univaq.it, piergiuseppe.dimarco@univaq.it,\\ fortunato.santucci@univaq.it, 
		\IEEEauthorrefmark{2}Email: aamir.mahmood@miun.se, mikael.gidlund@miun.se
}
\vspace{-24pt}
}

\maketitle

\begin{abstract}

Emerging 6G industrial IoT architectures require wireless networked control systems capable of stabilizing diverse control loops over tightly constrained radio resources. Conventional periodic and Age-of-Information (AoI) based scheduling guarantees bounded staleness at the cost of persistent channel saturation. Conversely, pure event-triggered (PureET) strategies minimize transmissions but risk catastrophic silent deterioration when local sensor-side thresholds fail to reflect critical state evolution. To bridge this gap, we propose a communication-control co-design framework governed by a 6G Semantic Layer that independently arbitrates uplink and downlink resources. Instead of relying on freshness, our architecture evaluates the actual control impact of a packet using the state-to-error ratio (SER). We unify this control confidence with channel reliability in terms of signal-to-noise ratio (SNR) to orchestrate a threshold-based sensor push and a state-aware controller pull mechanism. To ensure equitable resource allocation across dynamically heterogeneous plants, the proposed framework explicitly scales actuation deadbands according to local plant dynamics. Simulations over Rayleigh-faded channels demonstrate that this approach fundamentally shifts the Pareto frontier between transmission rate and control quality. The proposed scheme achieves tracking accuracy comparable to periodic schedulers at a reduced communication overhead, while mitigating the estimation errors characteristic of PureET.

\end{abstract}


% ==============================================================================
\section{Introduction}
% ==============================================================================
%\IEEEPARstart{I}{ndustrial-IoT} 
Industrial-IoT (IIoT) deployments and large-scale digital twins require stabilizing physical processes over shared wireless infrastructure~\cite{park2017wireless}. Decoupling wired loops into wireless networked control systems (WNCS) forces sensors and actuators to compete for finite radio resources, challenging 6G ultra-reliable low-latency communication (uRLLC) targets~\cite{8869705}. In dense deployments, control loops compete for scarce radio resources against diverse traffic, making intelligent resource allocation critical.

However, this scarcity reframes the scheduling problem: a finite resource pool must stabilize heterogeneous loops without causing network collapse. Conventional throughput-centric schedulers fail for control workloads because measurements from plants near equilibrium are highly predictable and largely redundant, while updates from disturbed plants are critical to prevent closed-loop instability. Bandwidth and information value rarely align, and schedulers blind to this mismatch waste resources while starving critical processes.

Periodic and Age of Information (AoI) approaches prioritize transmissions based on data staleness~\cite{yates2021age, ayan2019age}. While analytical and robust, freshness is only a proxy for relevance, and periodic schemes amplify this inefficiency by demanding full channel utilization to bound staleness regardless of instantaneous need. Pure event-triggered (PureET) architectures shift the decision to the sensor, transmitting only when locally observed innovations exceed a threshold~\cite{trimpe2014event, wang2010event}. This enables transmission suppression during steady-state operation but introduces a critical vulnerability. If triggering conditions miss relevant state evolution due to unobserved drift, process disturbances, or packet losses, the controller misses corrective updates and estimation degradation proceeds unchecked, risking loop destabilization. PureET and periodic scheduling thus represent opposing extremes: bandwidth-efficient but exposed to silent deterioration on one side, robust but resource-intensive on the other, revealing that information utility is inherently state-dependent.

A defining shift in recent 6G cyber-physical research is the move toward goal-oriented semantic communication-control co-design. 
Goal-oriented communication indicates a paradigm where transmission decisions are governed strictly by their contribution to the specific control objective, such as stabilizing an unstable plant, rather than merely fulfilling network-centric requirements. Complementary to this, the semantic approach refers to a framework where the value of information is quantified by its utility in reducing closed-loop estimation error, to prioritize updates that carry the highest functional relevance to the task at hand~\cite{gunduz2022beyond}. Recent frameworks operationalize this through AoI and Value of Information metrics~\cite{10618994, 10473131}, context-aware estimation~\cite{10731708, 11072837}, and joint digital twin orchestration~\cite{11311168, 11016699}. Unifying abstractions like the Semantic Aggregation Layer~\cite{ayan2024enabling} enable context-aware scheduling, yet limitations persist within these foundational models.

Existing semantic-scheduling frameworks frequently rely on assumptions that limit practical applicability. Most focus exclusively on the sensor-to-controller uplink, abstracting away controller-to-actuator downlink contention~\cite{ayan2019age}. This asymmetric modeling reflects a broader divide that isolates control confidence from channel reliability, missing a fundamental structural duality between them. To operationalize control confidence, we adopt the State-to-Error Ratio (SER), a metric used in networked control systems to quantify a controller's confidence by measuring the ratio of the estimated state magnitude to the estimation error. A trustworthy predictor (high SER) and a robust channel (high SNR) serve identical purposes: both guarantee sufficient quality to safely defer transmissions ~\cite{11027533}. Standard event-triggered architectures also struggle with heterogeneous plants, as uniform thresholds induce fairness violations, with high-gain control loops ignoring major drifts while low-gain loops overreact to minor noise. Resolving these intertwined challenges requires a unified architecture that jointly manages duplex capacity, exploits SER-SNR duality, and normalizes deadbands to individual dynamics.

Addressing these gaps, we propose a semantic-aware communication-control co-design framework, adopting semantic communication as a paradigm where transmission value is quantified exclusively by its capacity to reduce closed-loop estimation error. The framework bridges event-triggered efficiency at equilibrium with periodic robustness under stress, operationalizing the SER-SNR uniformity principle for broader theoretical impact. Our core technical contributions are fourfold:
\begin{itemize}
\item Arbitrate independent uplink and downlink capacity budgets under a unified semantic prioritization policy.
\item Encode control-side predictive confidence and communication-side channel reliability into a single, interpretable transmission-suppression rule.
\item Neutralize plant heterogeneity within the actuation deadband to guarantee equitable evaluation across diverse physical dynamics.
\item Counteract silent degradation in PureET paradigms via an AoI-driven pull-request mechanism empowering controllers to solicit updates when estimates deteriorate.
\end{itemize}

The remainder of this paper is organized as follows. Section~\ref{sec:system_model} presents the system model. Section~\ref{sec:semantic_metrics} defines the semantic metrics and estimation architecture. Section~\ref{sec:proposed_framework} details the scheduling and control framework. Section~\ref{sec:results} reports the simulation results, and Section~\ref{sec:conclusion} concludes this study.

% ==============================================================================
\section{System Model}
\label{sec:system_model}
\begin{figure}[!t]
    \centering
    \includegraphics[width=0.45\textwidth]{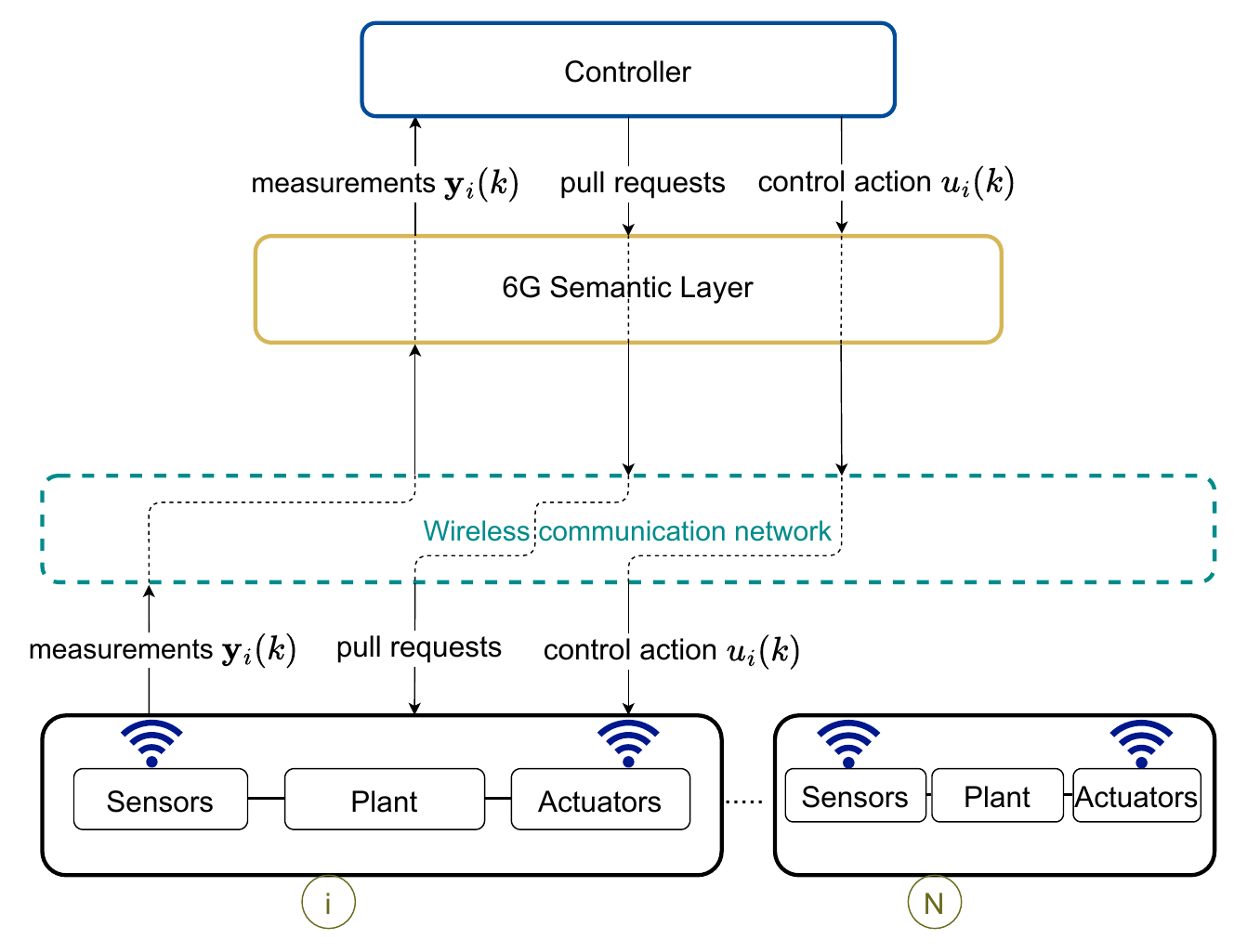}
    \vspace{-10pt}
    \caption{Semantic push–pull architecture for the proposed communication-control co-design framework. The 6G Semantic Layer coordinates sensor-initiated push updates, controller-initiated pull requests, and downlink control actions over shared wireless resources.}
    \vspace{-10pt}
    \label{fig:system_model}
\end{figure}

% ==============================================================================
We consider a WNCS where a central edge controller stabilizes $N$ spatially distributed unstable plants (e.g., robotic arms or automated guided vehicles) over a shared 5G/6G access point. Each plant is equipped with a local sensor and actuator that communicate with a centralized edge controller over bandwidth-constrained bidirectional wireless links. Uplink and downlink transmissions are allocated over orthogonal FDMA resource blocks, with separate capacity budgets for sensor updates and actuator commands. This setup reflects industrial IoT deployments where numerous high-rate control loops compete for limited radio resources. The architecture is depicted in Fig.~\ref{fig:system_model}.

The plants, indexed by $\mathcal{N}=\{1,2,\dots,N\}$, evolve in discrete time with a sampling period $T_s$ and index $k \in \{0,1,\dots,K-1\}$. At each step, the edge controller arbitrates limited uplink (UL) and downlink (DL) channels. The UL carries sensor-initiated \emph{push} updates (triggered by local innovation thresholds) or controller-initiated \emph{pull} requests (issued upon estimator degradation). The DL delivers resulting control commands to the actuators.

\subsection{Plant Dynamics}
Each plant $i \in \mathcal{N}$ is modeled as an inverted pendulum on a cart, a benchmark for networked control protocols due to its open-loop instability. The state vector $\mathbf{x}_i(k) = \begin{bmatrix} p_i(k) & \dot{p}_i(k) & \theta_i(k) & \dot{\theta}_i(k) \end{bmatrix}^{\!\top}$ captures position, velocity, pendulum angle, and angular velocity, respectively. Linearizing the continuous-time dynamics around the upright equilibrium and applying a zero-order hold at sampling period $T_s$ yields the discrete-time evolution,
\begin{equation}
\mathbf{x}_i(k+1) = \mathbf{A}_i \mathbf{x}_i(k) + \mathbf{B}_i u_i(k) + \mathbf{w}_i(k),
\label{eq:plant_dynamics}
\end{equation}
% where $\mathbf{A}_i \in \mathbb{R}^{4\times4}$ and $\mathbf{B}_i \in \mathbb{R}^{4\times1}$ denote the discrete-time state-transition and input matrices respectively, $u_i(k) \in \mathbb{R}$ is the horizontal control force and $\mathbf{w}_i(k) \sim \mathcal{N}(\mathbf{0}, \mathbf{Q}_w)$ models additive process disturbances with covariance $\mathbf{Q}_w \in \mathbb{R}^{4\times 4}$ characterizing per-axis noise intensity. Sensors observe the complete state through an additive Gaussian measurement model,
where $\mathbf{A}_i \in \mathbb{R}^{4\times4}$, $\mathbf{B}_i \in \mathbb{R}^{4\times1}$ are the state-transition and input matrices, $u_i(k) \in \mathbb{R}$ is the horizontal control force, and $\mathbf{w}_i(k) \sim \mathcal{N}(\mathbf{0}, \mathbf{Q}_w)$ models process disturbances with covariance $\mathbf{Q}_w \in \mathbb{R}^{4\times 4}$. Sensors observe the complete state via an additive Gaussian model,
\begin{equation}
    \mathbf{y}_i(k) = \mathbf{C}\,\mathbf{x}_i(k) + \mathbf{v}_i(k),
    \qquad \mathbf{v}_i(k) \sim \mathcal{N}(\mathbf{0}, \mathbf{R}_v),
    \label{eq:measurement}
\end{equation}
with $\mathbf{C} = \mathbf{I}_4$ and measurement noise covariance $\mathbf{R}_v$. The controller employs a discrete-time Linear Quadratic Regulator (LQR) law $\mathbf{K}_i = -\mathrm{dlqr}(\mathbf{A}_i,\mathbf{B}_i,\mathbf{Q},\mathbf{R})$, where $\mathbf{Q}$ and $\mathbf{R}$ are the weighting matrices penalizing state deviations and control effort, respectively, prioritizing angular stability. LQR is adopted for analytical tractability and clean separation of control and communication layers.
\subsection{Shared Wireless Network}
% At each step $k$, the scheduler grants UL access to at most $M_{\mathrm{UL}} \ll N$ plants and DL access to at most $M_{\mathrm{DL}} \ll N$ plants. Letting $\mathcal{G}^{\mathrm{UL}}(k), \mathcal{G}^{\mathrm{DL}}(k) \subseteq \mathcal{N}$ denote the scheduled sets,
At each step $k$, the scheduler grants UL/DL access to at most $M_{\mathrm{UL}}, M_{\mathrm{DL}} \ll N$ plants, respectively. Letting $\mathcal{G}^{\mathrm{UL}}(k), \mathcal{G}^{\mathrm{DL}}(k) \subseteq \mathcal{N}$ denote the scheduled sets,
\begin{equation}
|\mathcal{G}^{\mathrm{UL}}(k)| \le M_{\mathrm{UL}}, \qquad |\mathcal{G}^{\mathrm{DL}}(k)| \le M_{\mathrm{DL}}.
\label{eq:channel_cap}
\end{equation}
Channel impairment is isolated to the UL, reflecting power-constrained sensor nodes in IIoT. The UL is modeled as a Rayleigh block-fading channel with outage-based reception. The instantaneous received SNR is $\gamma_i(k) = \bar{\gamma}\,\big|h_i(k)\big|^2$, yielding the closed-form packet error rate $P_{\mathrm{out}} = \Pr\!\left[\gamma < \gamma_{\mathrm{th}}\right] = 1 - \exp\!\left(-\gamma_{\mathrm{th}}/\bar{\gamma}\right)$, where $\gamma_{\mathrm{th}}$ is the minimum decoding threshold. With $S_i(k) \in \{0,1\}$ denoting the success indicator, the effective reception sets are
\begin{equation}
\mathcal{R}^{\mathrm{UL}}(k) = \left\{ i \in \mathcal{G}^{\mathrm{UL}}(k) : S_i(k) = 1 \right\},
\label{eq:effective_rx_ul}
\end{equation}
while the downlink is treated as capacity-limited but error-free,
\begin{equation}
\mathcal{R}^{\mathrm{DL}}(k) = \mathcal{G}^{\mathrm{DL}}(k).
\label{eq:effective_rx_dl}
\end{equation}
This asymmetric assumption reflects cell-edge IIoT realities, where the base station enjoys higher transmit power and richer antenna provisioning than the sensor nodes. In our framework, the sensor only checks local innovation, while the controller or 6G Semantic Layer (6G~SL) uses channel-quality information for scheduling.

\section{Semantic Metrics and Estimation Architecture}
\label{sec:semantic_metrics}

Translating raw physical signals into semantic scheduling decisions requires tracking data freshness and predictive confidence. This section defines the core metrics and estimation architecture supporting our framework.

\subsection{Information Freshness}
Freshness is tracked via the AoI abstraction. Two separate AoI counters are maintained per plant
\begin{align}
    \Delta_i^{\mathrm{c}}(k+1) &=
        \begin{cases}
            1, & i \in \mathcal{R}^{\mathrm{UL}}(k), \\[2pt]
            \Delta_i^{\mathrm{c}}(k) + 1, & \text{otherwise},
        \end{cases}
        \label{eq:aoi_sensor}\\[4pt]
    \Delta_i^{\mathrm{a}}(k+1) &=
        \begin{cases}
            1, & i \in \mathcal{R}^{\mathrm{DL}}(k), \\[2pt]
            \Delta_i^{\mathrm{a}}(k) + 1, & \text{otherwise}.
        \end{cases}
        \label{eq:aoi_control}
\end{align}

The controller-side AoI $\Delta_i^{\mathrm{c}}(k)$ quantifies the time since the controller last received a fresh measurement, while the actuator-side AoI $\Delta_i^{\mathrm{a}}(k)$ measures actuator staleness. The proposed scheduler exploits this two-sided view to gate pull requests and downlink priority boosts.

% The sensor-side AoI $\Delta_i^{\mathrm{s}}(k)$ quantifies the time since the last fresh measurement, while the control-side AoI $\Delta_i^{\mathrm{c}}(k)$ measures actuator staleness. The proposed scheduler exploits this two-sided view to gate pull requests and downlink priority boosts.

\subsection{Controller-Side Predictor}
% Between measurement arrivals, the edge controller maintains a model-based predictor for each plant, implemented as a two-stage measurement-update/propagation cycle. With $\hat{\mathbf{x}}_i(k|k)$ denoting the filtered estimate at step $k$ (incorporating all data available up to and including step $k$) and $\hat{\mathbf{x}}_i(k|k-1)$ the predicted estimate at step $k$ using data only up to step $k-1$,
The edge controller maintains a two-stage update/propagation predictor for each plant. With $\hat{\mathbf{x}}_i(k|k)$ the filtered estimate at step $k$ (using data up to step $k$) and $\hat{\mathbf{x}}_i(k|k-1)$ the predicted estimate at step $k$ (using data only up to $k-1$),
\begin{subequations}
\label{eq:predictor}
\begin{align}
\hat{\mathbf{x}}_i(k|k) &=
\begin{cases}
\mathbf{y}_i(k), & i \in \mathcal{R}^{\mathrm{UL}}(k), \\
\hat{\mathbf{x}}_i(k|k-1), & \text{otherwise},
\end{cases}
\label{eq:predictor_update} \\[4pt]
\hat{\mathbf{x}}_i(k+1|k) &= \mathbf{A}_i\hat{\mathbf{x}}_i(k|k) + \mathbf{B}_i u_i(k).
\label{eq:predictor_predict}
\end{align}
\end{subequations}
Eq.~\eqref{eq:predictor_update} updates the estimate upon a successful uplink or retains the prior prediction otherwise, while~\eqref{eq:predictor_predict} propagates it through the plant dynamics. As $\Delta_i^{\mathrm{c}}(k)$ increases, accumulated noise and linearization error degrade $\hat{\mathbf{x}}_i$, a drift that triggers controller pull requests.

% Between measurement arrivals, the edge controller maintains a model-based predictor for each plant. With $\hat{\mathbf{x}}_i(k)$ denoting the estimated state at step $k$
% \begin{equation}
% \hat{\mathbf{x}}_i(k+1) =
% \begin{cases}
% \mathbf{y}_i(k), & i \in \mathcal{R}^{\mathrm{UL}}(k), \\[3pt]
% \mathbf{A}_i\,\hat{\mathbf{x}}_i(k) + \mathbf{B}_i\,u_i(k), & \text{otherwise}.
% \end{cases}
% \label{eq:predictor}
% \end{equation}
% Although written as a single recursion,~\eqref{eq:predictor} captures a two-stage process analogous to standard filtering. A successful uplink at step $k$ instantly resets the estimate using the fresh measurement $\mathbf{y}_i(k)$ before the dynamics roll forward, while a failed or absent uplink forces open-loop extrapolation from the prior estimate. Estimation quality degrades with $\Delta_i^{\mathrm{s}}(k)$, as accumulated process noise and linearization error push $\hat{\mathbf{x}}_i$ away from the true $\mathbf{x}_i$, a drift we utilize to trigger controller-initiated pull requests.

\subsection{State-to-Error Ratio: A Semantic Confidence Metric}

When a fresh measurement $\mathbf{y}_i(k)$ arrives, comparing it against the pre-update prediction $\hat{\mathbf{x}}_i^-(k) := \mathbf{A}_i \hat{\mathbf{x}}_i(k-1) + \mathbf{B}_i u_i(k-1)$ yields a local indicator of predictor fidelity. We define the instantaneous State-to-Error Ratio (SER) as
\begin{equation}
    \mathrm{SER}_i(k) \;=\; \frac{\|\hat{\mathbf{x}}_i^-(k)\|^2 + \epsilon}{\|\mathbf{y}_i(k) - \hat{\mathbf{x}}_i^-(k)\|^2 + \epsilon},
    \label{eq:ser_raw}
\end{equation}
with $\epsilon > 0$ a small regularizer ensuring the ratio remains well-defined near equilibrium. A large $\mathrm{SER}_i$ signals that the innovation $\|\mathbf{y}_i - \hat{\mathbf{x}}_i^-\|$ is small relative to the estimated state, so the predictor can be trusted, while a small value indicates open-loop extrapolation is unsafe. Since no innovation is available between measurements, we propagate the last computed $\mathrm{SER}_i$ with a monotone decay tied to sensor AoI as
\begin{equation}
    \mathrm{SER}^{\mathrm{eff}}_i(k) \!=\! \min\!\left(\mathrm{SER}_{\max},\; \frac{\mathrm{SER}_i}{\min(\Delta_i^{\mathrm{c}}(k),\,\Delta_{\mathrm{decay}})}\right).
    \label{eq:ser_eff}
\end{equation}
The saturation floor $\Delta_{\mathrm{decay}}$ prevents unbounded decay during prolonged outages, and $\mathrm{SER}_{\max}$ caps the ratio near equilibrium. The effective SER in~\eqref{eq:ser_eff} serves as the control-side semantic confidence signal.

\subsection{Performance Objective}
Our design goal is to stabilize all plants while minimizing the system-wide number of transmissions. The \emph{control cost}, expressed as the time-averaged root-mean-square (RMS) angular deviation
\begin{equation}
    J_{\theta}^{\mathrm{RMS}} = \sqrt{\frac{1}{KN}\sum_{k=0}^{K-1}\sum_{i=1}^{N}\theta_i^2(k)},
    \label{eq:obj_theta}
\end{equation}
penalizes poor stabilization. The \emph{communication cost}, defined as the total cumulative count of successful uplink transmissions across the network over the time horizon $K$,
\begin{equation}
    J_{\mathrm{tx}} = \sum_{k=0}^{K-1}\sum_{i=1}^{N}\mathds{1}\!\left\{i \in \mathcal{R}^{\mathrm{UL}}(k)\right\},
    \label{eq:obj_tx}
\end{equation}
penalizes wasteful bandwidth use. A preferable scheduler minimizes both quantities simultaneously.

% ==============================================================================
\section{Proposed Framework for Semantic-Aware Scheduling and Control}
\label{sec:proposed_framework}
% ==============================================================================
Our proposed framework builds on a single organizing principle: a channel slot is worth spending only when the transmission's content meaningfully alters the closed-loop outcome. We implement this via decentralized event-triggering, 6G~SL arbitration, and edge controller logic.

\subsection{Decentralized Semantic Event-Triggering at the Sensor}

Each sensor $i$ maintains a local memory of its last successfully transmitted measurement $\big(\mathbf{y}_i^{\mathrm{last}},\,\theta_i^{\mathrm{last}}\big)$. It evaluates a local innovation score $\eta_i^{\mathrm{s}}(k) = \big|\theta_i(k) - \theta_i^{\mathrm{last}}\big|$, and generates an event-driven measurement data unit (MDU) only when $\eta_i^{\mathrm{s}}(k) > \delta^{\mathrm{s}}_i$. The threshold $\delta^{\mathrm{s}}_i$ is tailored to plant dynamics. When the trigger fires, the sensor submits a \emph{push} MDU carrying the current measurement, its generation age $\Delta_i^{\mathrm{c}}(k)$, and a priority
\begin{equation}
    \rho_i^{\mathrm{push}}(k) = \frac{\eta_i^{\mathrm{s}}(k)}{\delta^{\mathrm{s}}_i},
    \label{eq:push_prio}
\end{equation}
% \begin{equation}
%     \rho_i^{\mathrm{push}}(k) = \frac{\delta_i^{\mathrm{s}}(k)}{\Delta^{\mathrm{s}}_i},
%     \label{eq:push_prio}
% \end{equation}
which normalizes urgency across plant types. Silent sensors contribute no traffic. The locality is important: the sensor never needs to know the global plant population, channel state, or scheduler queue, which makes the trigger implementable at the extreme edge of the network.

\subsection{6G Semantic Layer and Duplex Arbitration}
% We propose the 6G SL between the application logic and radio resource scheduler, aggregating events, prioritizing semantically, and arbitrating duplex links. 

We propose the 6G SL between the application logic and radio resource scheduler as the AI-native semantic arbitration function envisioned for 6G RAN architectures~\cite{8869705}, rather than a generic traffic scheduler retrofitted with priority tags. Every communication event carries semantic metadata (type, priority, age-at-generation) that the scheduler can act upon, and the layer plays three roles: 
\begin{itemize}
    \item \emph{Event aggregation:} merging sensor-side push events and controller-side pull requests into a single priority queue.
    \item \emph{Semantic prioritization:} ranking queue entries by a composite score that combines AoI, innovation magnitude $\|\mathbf{y}_i(k) - \hat{\mathbf{x}}_i(k)\|$ (the discrepancy between a fresh measurement and the predictor's estimate, capturing how much new information the update carries), and downlink actuator staleness.
    \item \emph{Duplex arbitration:} granting the top $M_{\mathrm{UL}}$ uplink entries and the top $M_{\mathrm{DL}}$ pending control updates independently within their capacity budgets.
\end{itemize}

The uplink queue $\mathcal{Q}^{\mathrm{UL}}(k)$ aggregates two sources: sensor-originated push MDUs that cleared their local triggers at step $k$, and controller-initiated pull MDUs corresponding to plants whose pull flag $\pi_i(k)$ has been raised by the \textit{Condition~C} in Sec.~\ref{subsec:controller_logic}. Each pull inherits a priority that scales with controller-side AoI as
\begin{equation}
\rho_i^{\mathrm{pull}}(k) = \frac{\Delta_i^{\mathrm{c}}(k)}{\max\!\left(\tau^{\mathrm{c}},\,1\right)},
\label{eq:pull_prio}
\end{equation}
escalating monotonically as the estimate ages. The DL queue $\mathcal{Q}^{\mathrm{DL}}(k)$ collects control actions carrying a base priority $\rho_i^{\mathrm{base}}(k)$ proportional to the planned change magnitude, with a staleness-aware boost keeping long-frozen actuators competitive even when their raw priority is modest
\begin{equation}
\rho_i^{\mathrm{DL}}(k) = \rho_i^{\mathrm{base}}(k) + \beta\,\mathds{1}\!\left\{\Delta_i^{\mathrm{a}}(k) \ge \tau^{\mathrm{a}}\right\}\,\frac{\Delta_i^{\mathrm{a}}(k)}{\tau^{\mathrm{a}}},
\label{eq:dl_prio}
\end{equation}
with $\beta > 0$ controlling the boost strength.

Both queues are sorted in descending priority and deduplicated, ensuring that no plant occupies more than one slot per direction in a given step. Letting $\tilde{\mathcal{Q}}^{\mathrm{UL}}(k)$ and $\tilde{\mathcal{Q}}^{\mathrm{DL}}(k)$ denote the resulting sorted lists, the grant sets follow as
\begin{equation}
\begin{split}
\mathcal{G}^{\mathrm{UL}}(k) &= \mathrm{TopM}\!\left(\tilde{\mathcal{Q}}^{\mathrm{UL}}(k),\,M_{\mathrm{UL}}\right), \\
\mathcal{G}^{\mathrm{DL}}(k) &= \mathrm{TopM}\!\left(\tilde{\mathcal{Q}}^{\mathrm{DL}}(k),\,M_{\mathrm{DL}}\right),
\end{split}
\label{eq:duplex_grant}
\end{equation}
where $\mathrm{TopM}(\cdot,m)$ extracts the first $m$ elements. When fewer than $M_{\mathrm{UL}}$ uplink entries are queued, the remaining slots are deliberately left \emph{idle}. The 6G SL never fabricates traffic to fill the channel, operationalizing the principle that silence itself carries information. Under the uplink-only fading model, downlink winners receive their commands reliably; capacity truncation in~\eqref{eq:duplex_grant} is the sole source of downlink failure. Plants outside $\mathcal{G}^{\mathrm{DL}}(k)$ retain their previous input.
% The 6G SL never fabricates traffic to fill the channel, operationalizing the semantic principle that silence itself carries information. Under the uplink-only fading model adopted here, downlink winners have their commands delivered reliably, with capacity contention already captured by the truncation in~\eqref{eq:duplex_grant} being the only source of downlink failure. Plants outside $\mathcal{G}^{\mathrm{DL}}(k)$ retain their previously applied input.

\subsection{Edge Controller Logic}
\label{subsec:controller_logic}
The edge controller executes once per slot on a per-plant basis, partitioning every (plant, slot) pair into one of three mutually exclusive branches (formalized in Algorithm~\ref{alg:controller}):

\begin{algorithm}[htbp]
\caption{Per-slot edge controller logic for plant $i$}
\label{alg:controller}
\begin{algorithmic}[1]
\REQUIRE $\mathcal{R}^{\mathrm{UL}}(k)$, $\mathbf{y}_i(k)$, $\hat{\mathbf{x}}_i(k)$, $u_i(k)$, $\Delta_i^{\mathrm{c}}(k)$, $\mathrm{SER}^{\mathrm{eff}}_i(k)$, $\gamma_i^{\mathrm{UL}}(k)$
\STATE \textbf{Constants:} $\mathrm{SER}_{\mathrm{th}}$, $\gamma_{\mathrm{th}}$, $\lambda$, $\tau^{\mathrm{c}}$, $\Delta\theta_{\mathrm{base}}$
\IF{$i \in \mathcal{R}^{\mathrm{UL}}(k)$}
    \STATE \textit{\textbf{Condition A:} fresh measurement}
    \STATE $\hat{\mathbf{x}}_i(k) \gets \mathbf{y}_i(k)$
    \STATE $u_i^{\star}(k) \gets \mathbf{K}_i\,\hat{\mathbf{x}}_i(k)$
    \STATE $\rho_i^{\mathrm{base}}(k) \gets |u_i^{\star}(k) - u_i(k)| + 1.0$
    \STATE Queue control action with priority $\rho_i^{\mathrm{base}}(k)$
    \STATE $\pi_i(k) \gets 0$
\ELSE
    \STATE $U_i(k) \gets \frac{\mathrm{SER}^{\mathrm{eff}}_i(k)}{\mathrm{SER}_{\mathrm{th}}} + \lambda\,\frac{\gamma_{\mathrm{th}}}{\gamma_i^{\mathrm{UL}}(k)}$
    \IF{$U_i(k) \ge 1$}
        \STATE \textit{\textbf{Condition B:} silence}
        \STATE \text{no action queued}
    \ELSE
        \STATE \textit{\textbf{Condition C:} stale estimate}
        \STATE $u_i^{\star}(k) \gets \mathbf{K}_i\,\hat{\mathbf{x}}_i(k)$
        \STATE $\Delta^u_i \gets |K_{i,3}| \cdot \Delta\theta_{\mathrm{base}}$
        \IF{$|u_i^{\star}(k) - u_i(k)| > \Delta^u_i$}
            \STATE Queue control action with priority $|u_i^{\star}(k) - u_i(k)|$
        \ENDIF
        \STATE $\pi_i(k) \gets \mathds{1}\{\Delta_i^{\mathrm{c}}(k) \ge \tau^{\mathrm{c}}\}$
    \ENDIF
\ENDIF
\end{algorithmic}
\end{algorithm}

\begin{itemize}
\item \emph{Condition A (Fresh Measurement):} When $i \in \mathcal{R}^{\mathrm{UL}}(k)$, the predictor synchronizes, generating a fresh command queued with high priority, clearing the pull flag.
\item \emph{Condition B (Channel-Aware Silence):} The controller evaluates a semantic uniformity score combining predictive confidence ($\mathrm{SER}$) and channel quality ($\gamma^{\mathrm{UL}}$). This uniformity score integrates SER and SNR as complementary indicators governing transmission suppression. A high SER directly enforces silence by confirming predictor reliability, rendering fresh updates semantically unnecessary. The SNR term functions as a conditional penalty: a degraded link inflates the score to block likely-to-fail transmissions, whereas a strong channel neutralizes this penalty. A high SNR does not inherently force a transmission; it merely allows the scheduler to update the plant if the control side requires correction. Ultimately, these metrics suppress low-yield traffic across two distinct axes: SER evaluates whether new information is \emph{needed}, while SNR dictates whether it can be reliably \emph{delivered}.
\item \emph{Condition C (Stale Estimate):} Operating on open-loop extrapolation, the controller tests candidate commands against an angle-equivalent deadband $\Delta^u_i = |K_{i,3}|\cdot\Delta\theta_{\mathrm{base}}$. Negligible updates are discarded. If controller-side AoI exceeds threshold $\tau^{\mathrm{c}}$, the pull flag $\pi_i$ is raised to solicit fresh data.
\end{itemize}

\section{Results and Discussion}
\label{sec:results}

We evaluate the proposed framework against four baselines on a heterogeneous population of $N=30$ inverted pendulums sharing a duplex-constrained 6G uplink under Rayleigh block fading. To capture diverse control requirements, the plants are evenly cycled across three distinct instability classes: highly unstable ($L=0.2$\,m), moderately unstable ($L=0.5$\,m), and slightly unstable ($L=1.0$\,m) dynamics. Sharing $M_{\mathrm{UL}}=10$ resource blocks among these 30 plants establishes a strict $3:1$ contention ratio, explicitly chosen to stress the network and represent a congested industrial workcell. The proposed architecture scales efficiently to larger populations due to its asymmetric design: sensor-side complexity remains strictly $\mathcal{O}(1)$, offloading the $\mathcal{O}(N)$ uniformity evaluations to the resource-rich edge controller. Furthermore, as $N$ grows, the SER-SNR rule inherently suppresses excess traffic, gracefully degrading individual update rates to prevent total channel collapse under extreme plant densities. The evaluated baselines (sharing identical physical and channel parameters) are configured as follows:
\begin{itemize}
    \item \emph{PureET}: transmits only when the local angular innovation crosses the per-plant threshold $\delta^{\mathrm{s}}_i$, with no controller-initiated pulls and no SER-SNR uniformity test.
    \item \emph{Round-Robin}: cyclically grants the $M_{\mathrm{UL}}$ uplink slots to plants in fixed order, ignoring plant state and channel quality. 
    \item \emph{Max-AoI}: greedily allocates each slot to the $M_{\mathrm{UL}}$ plants with the largest controller-side AoI $\Delta_i^{\mathrm{c}}(k)$, prioritizing freshness without semantic context.
    \item \emph{Random}: selects $M_{\mathrm{UL}}$ plants uniformly at random per slot, serving as a stateless reference. 
\end{itemize}
Table~\ref{tab:sim_results} consolidates the simulation configuration and the per-method results.

\begin{table}[t]
\caption{Simulation parameters and per-method performance.\\ Evaluated metrics: $u_{\mathrm{Energy}}{=} \sum u^2 T_s$ the cumulative actuation energy; $J_{\mathrm{cc}}{=}(J_{\mathrm{tx}}^{\mathrm{UL}}{+}J_{\mathrm{tx}}^{\mathrm{DL}})\cdot(\theta_{\mathrm{RMS}})^2$ the joint communication-control cost; Jain {=} per-segment Jain fairness index over 1\,s windows.}
\label{tab:sim_results}
\centering
\renewcommand{\arraystretch}{1.05}
\setlength{\tabcolsep}{3pt}
\footnotesize
\resizebox{\columnwidth}{!}{%
\begin{tabular}{@{}lcccc@{}}
\toprule
\multicolumn{5}{c}{\textbf{Configuration:} $N{=}30$, $T_s{=}10$~ms, $T_{\mathrm{sim}}{=}60$~s,$\lambda{=}1.0$, $\mathrm{SER}_{\mathrm{th}}{=}50$, $\gamma_{\mathrm{th}}{=}0$~dB} \\
\multicolumn{5}{c}{$M_{\mathrm{UL}}{=}M_{\mathrm{DL}}{=}10$, $P_{\mathrm{out}} \in [0.05, 0.50]$, $\bar{\gamma}{=}9.77$~dB, $\tau^{\mathrm{c}}{=}\tau^{\mathrm{a}}{=}10$, $\beta{=}10$, $\Delta\theta_{\mathrm{base}}{=}0.3^\circ$} \\
\midrule
\textbf{Method} & $\theta_{\mathrm{RMS}}$ (deg) & $\theta_{\max}$ (deg) & $u_{\mathrm{Energy}}$ (N$^2$s) & $J_{\mathrm{cc}}$ \\
\midrule
\textbf{Proposed (6G SL)} & \textbf{0.80} & \textbf{6.02} & \textbf{484}   & \textbf{2969}  \\
PureET                    & 1.95          & 9.56          & 28867          & 43728 \\
Round-Robin               & 0.49          & 6.62          & 188            & 10728  \\
Max-AoI                   & 0.47          & 5.93          & 150            & 9881  \\
Random                    & 0.59          & 14.67         & 816            & 17280  \\
\midrule
\textbf{Method} & UL Util. (\%) & UL Tx (pkts) & DL Tx (pkts) & Jain (1\,s) \\
\midrule
\textbf{Proposed (6G SL)} & \textbf{7.7} & \textbf{4610}  & \textbf{21369} & \textbf{0.967} \\
PureET                    & 19.1          & 11443          & 11443 & 0.982 \\
Round-Robin               & 75.7          & 45445          & 45445 & 1.000 \\
Max-AoI                   & 75.7          & 45445          & 45445 & 1.000 \\
Random                    & 75.7          & 45432          & 45432 & 0.999 \\
\bottomrule
\end{tabular}%
}
\end{table}

\begin{figure}[t]
\centering
\includegraphics[width=\columnwidth]{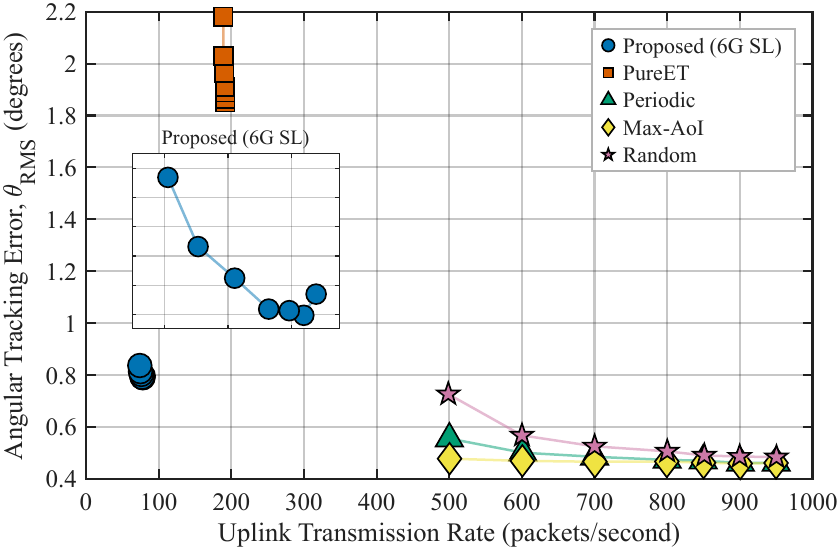}
\vspace{-18pt}
\caption{Empirical Pareto frontier in the (transmission rate, angular tracking error) plane. Operating points in the lower-left dominate.}
\vspace{-14pt}
\label{fig:pareto}
\end{figure}

Fig.~\ref{fig:pareto} maps each scheduler onto the bandwidth-versus-quality plane. Round-Robin and Max-AoI cluster in the high-bandwidth region at near-saturated channel use ($75\%$ UL utilization, $\sim$$757$~packets/s on the UL) and deliver the lowest $\theta_{\mathrm{RMS}}$ values ($0.49^\circ$, $0.47^\circ$), but only because every plant is being served whether or not it needs attention. PureET pushes toward opposite extreme, transmitting at $\sim$$191$~packets/s but paying a substantial control penalty ($\theta_{\mathrm{RMS}}{=}1.95^\circ$, peak $9.56^\circ$). Random collapses entirely under the heterogeneous-plant configuration, with peak excursions exceeding $14.67^\circ$ on the worst-served loops, confirming that stateless allocation cannot meet the demands of unstable dynamics. The proposed scheme uniquely occupies the lower-left region ($\theta_{\mathrm{RMS}}{=}0.80^\circ$at$\sim$$77$~packets/s and $7.7\%$ UL utilization), cutting UL traffic by $9.9\times$ over periodic baselines while staying within $\sim$$0.33^\circ$ of their tracking accuracy.
% The proposed scheme is the only point that lies genuinely lower-left: $\theta_{\mathrm{RMS}}{=}0.84^\circ$ at $\sim$$67$~packets/s on the UL with only $6.7\%$ UL utilization, a $7.5\times$ reduction in UL traffic relative to the periodic baselines while remaining within $\sim$$0.36^\circ$ of their tracking accuracy.

\begin{figure}[t]
\centering
\includegraphics[width=\columnwidth]{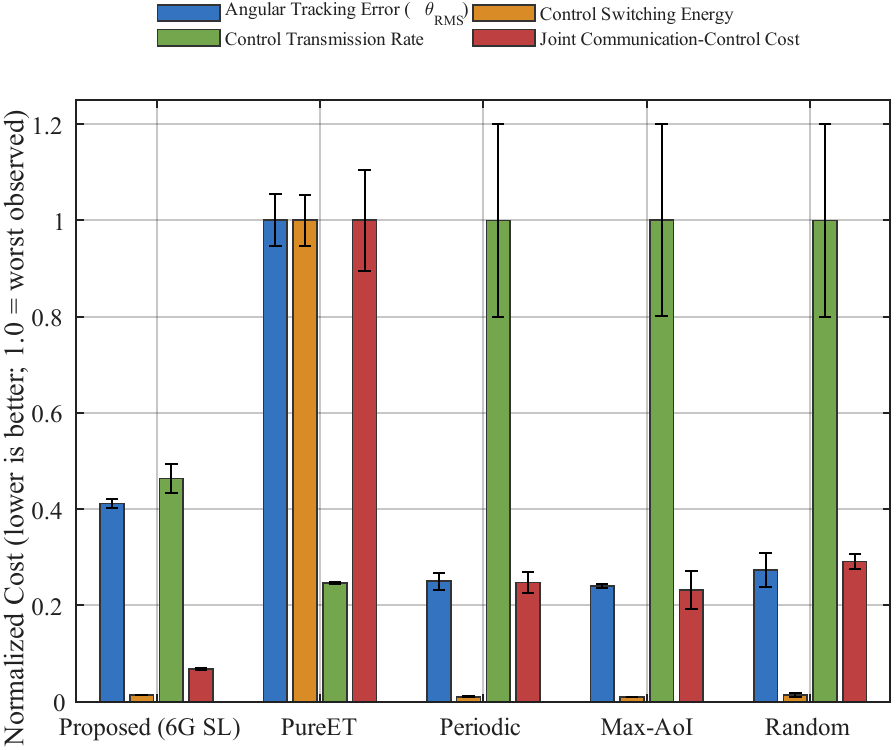}
\vspace{-16pt}
\caption{Normalized strategy comparison across four cost dimensions: Angular RMS Error, Actuation Switching Energy, Control Transmission Rate, and Joint Communication-Control Cost. Bars are normalized within each metric to the worst-performing method.}
\vspace{-12pt}
\label{fig:bar_chart}
\end{figure}

Fig.~\ref{fig:bar_chart} consolidates the comparison across four normalized cost dimensions. The proposed scheme is the only method whose bars stay below $\sim$$0.7$ on every axis simultaneously. PureET wins on the transmission-rate axis but its switching energy $u_{Energy}$ exceeds the proposed scheme's by $60\times$, exposing how scarce transmissions translate into large corrective inputs after silent drift. The proposed scheme's actuation energy reflects the per-plant gain-normalized deadband in Table~\ref{tab:sim_results}: aggressively-gained plants are not over-actuated on noise, and softly-gained plants do not require large remedial inputs because they are correctly tracked. The joint communication-control cost places the proposed scheme roughly $14.7\times$ below PureET, $5.8\times$ below Random, and $3.3$--$3.6\times$ below the periodic baselines, despite using a fraction of their bandwidth.

% The joint communication-control cost places the proposed scheme an order of magnitude below PureET and Random, and within a factor of $2$ of the periodic baselines despite using a fraction of their bandwidth.

Condition~C accounts for most transmission suppressions, showing that the framework primarily silences transmissions when control utility of a fresh update is negligible (falling within actuation deadband $\Delta^u_i$), rather than due to poor link quality. This behavior matches the evaluated high-SNR regime ($\bar{\gamma}{=}9.77$~dB), where channel penalties only dictate silence during rare outages. Lower SNRs naturally shift suppression toward Condition~B. Thus, the uniformity rule dynamically restricts traffic according to the active bottleneck, balancing control confidence and channel reliability. Concentrating bandwidth on active transients slightly lowers Jain's fairness index to $0.967$ (vs. $1.0$ for baselines) without starving plants. Through Pareto dominance, the framework absorbs stochastic channel impairments into its transmission–silence decisions, preventing closed-loop control failures.

% A source-aware breakdown of the simulation slots reveals that Condition~C accounts for the majority of transmission suppressions. This indicates that the framework primarily silences the channel when the control utility of a fresh update is negligible (falling within the actuation deadband $\Delta^u_i$), rather than in response to poor link quality. This behavior is expected given the strong average SNR ($\bar{\gamma}{=}9.77$~dB) of this evaluation, where the channel-cost penalty only dictates silence during rare outages. At lower SNRs, the locus of suppression naturally shifts toward Condition~B. Ultimately, this confirms that the uniformity rule dynamically restricts traffic based on the active bottleneck, smoothly pivoting between control-side confidence and channel-side reliability. As a safety net, the pull mechanism activates only when controller-side AoI exceeds $\tau^{\mathrm{c}}$. Concentrating bandwidth dynamically on active transients predictably lowers Jain's fairness index to $0.960$ relative to the baselines' near-perfect $1.0$, yet this deliberate allocation tradeoff avoids chronic plant starvation. Ultimately, by maintaining a Pareto-dominant profile, the framework successfully absorbs stochastic channel impairments into the transmission-silence balance, preventing them from propagating into closed-loop control failures.

% ==============================================================================
\section{Conclusion}\label{sec:conclusion}
% ==============================================================================
We presented a semantic-aware communication-control co-design framework for 6G WNCS, governed by a 6G Semantic Layer that arbitrates uplink and downlink resources independently, encodes SER and SNR into a unified silence rule, neutralizes plant heterogeneity through a gain-normalized actuation deadband, and closes the silent-deterioration loop of pure event-triggering via an AoI-driven controller-initiated pull. Across a heterogeneous population of inverted pendulums under Rayleigh-faded uplinks, the framework approaches the tracking accuracy of periodic schedulers at a fraction of their transmission rate while eliminating the divergent estimation excursions characteristic of PureET. Future work will extend the framework along three directions: formal stability and convergence guarantees under stochastic outages, downlink fading with PHY-layer link adaptation, and a learning-assisted Model Predictive Control layer that integrates cleanly above the existing communication substrate without altering the scheduling logic.

\section*{Acknowledgments}
This work is supported by the Swedish Knowledge Foundation via the IRS TransTech, Mid Sweden University, and by the Center of Excellence Ex-Emerge at the University of L'Aquila, funded by the Italian Government under CIPE Resolution 70/2017.

\bibliographystyle{IEEEtran}
\vspace{-2pt}
\bibliography{References}

\end{document}